\documentclass[conference]{IEEEtran}
\IEEEoverridecommandlockouts
\usepackage{cite}
\usepackage{amsmath,amssymb,amsfonts}
\usepackage{algorithmic}
\usepackage{graphicx}
\usepackage{svg}
\usepackage{textcomp}
\usepackage{xcolor}
\usepackage{multirow}
\usepackage{booktabs}
\usepackage{url}
\usepackage{booktabs}
\def\BibTeX{{\rm B\kern-.05em{\sc i\kern-.025em b}\kern-.08em
    T\kern-.1667em\lower.7ex\hbox{E}\kern-.125emX}}
\begin{document}

\newcommand{\TinyEquation}{\fontsize{6.5pt}{\baselineskip}\selectfont}
\newcommand{\SmallEquation}{\fontsize{7.0pt}{\baselineskip}\selectfont}
\newcommand{\MiddleEquation}{\fontsize{9pt}{\baselineskip}\selectfont}
\newcommand{\BigEquation}{\fontsize{9.5pt}{\baselineskip}\selectfont}
\newcommand{\ourmethod}{\textsc{WGAN-AFL}}

\title{\ourmethod{}: Seed Generation Augmented Fuzzer with Wasserstein-GAN}

\author{
  Liqun Yang\textsuperscript{\rm 1}, 
  Chunan Li\textsuperscript{\rm 1},
  Yongxin Qiu\textsuperscript{\rm 1}, 
  Chaoren Wei\textsuperscript{\rm 1}, 
  {Jian Yang}\textsuperscript{\rm 2}*, \\
  {Hongcheng Guo}\textsuperscript{\rm 2}, 
  {Jinxin Ma}\textsuperscript{\rm 3}, 
  {Zhoujun Li}\textsuperscript{\rm 2 }\\
  \textsuperscript{\rm 1}School of Cyber Science and Technology, Beihang University \\ 
  \textsuperscript{\rm 2}School of Computer Science and Engineering;  \\
  \textsuperscript{\rm 3}China Information Technology Security Evaluation Center; \\
  \{lqyang, lcn142857, 19377012, weichaoren, jiaya, hongchengguo, lizj\}@buaa.edu.cn; \\ 
  majinxin2003@126.com;
}


\maketitle

\begin{abstract}
The importance of addressing security vulnerabilities is indisputable, with software becoming crucial in sectors such as national defense and finance. Consequently, The security issues caused by software vulnerabilities cannot be ignored. Fuzz testing is an automated software testing technology that can detect vulnerabilities in the software. However, most previous fuzzers encounter challenges that fuzzing performance is sensitive to initial input seeds. In the absence of high-quality initial input seeds, fuzzers may expend significant resources on program path exploration, leading to a substantial decrease in the efficiency of vulnerability detection. To address this issue, we propose WGAN-AFL. By collecting high-quality testcases, we train a generative adversarial network (GAN) to learn their features, thereby obtaining high-quality initial input seeds. To overcome drawbacks like mode collapse and training instability inherent in GANs, we utilize the Wasserstein GAN (WGAN) architecture for training, further enhancing the quality of the generated seeds. Experimental results demonstrate that WGAN-AFL significantly outperforms the original AFL in terms of code coverage, new paths, and vulnerability discovery, demonstrating the effective enhancement of seed quality by WGAN-AFL.
\end{abstract}


\section{INTRODUCTION}
In the 21st century, software plays an irreplaceable role in various critical domains such as national defense, finance and economy \cite{DBLP:journals/cybersec/LiZZ18}. Consequently, the security issues caused by software vulnerabilities cannot be ignored. A vulnerability of high risk has the potential to enable remote manipulation of equipment, thereby precipitating the leakage of sensitive information and service interruptions. In more severe instances, it leads to the complete destabilization of the system, resulting in incalculable losses. A notable example is the 2016 incident where a Japanese satellite disintegrated due to an underlying software malfunction, incurring billions of dollars in losses \cite{news0}. Similarly, in 2018 and 2019, two fatal crashes involving Boeing 737MAX aircraft were attributed to software design defects, resulting in the tragic loss of 346 lives \cite{news1}.  Therefore, software needs to undergo rigorous testing and inspection to prevent irreversible losses caused by its vulnerabilities.

To effectively ensure software security, we can try to discover potential vulnerabilities before the software is put into use. Fuzz testing, as an automated software testing technology, generates random input that is unexpected by the target. Through monitoring the resulting abnormal results, such as software crashes, memory errors, etc., we can identify existing vulnerabilities in the software. Fuzz testing effectively reduces the human and time costs required for vulnerability discovery, making it possible for technical personnel who cannot discover vulnerabilities to identify vulnerabilities in the software through fuzz testing. Based on these advantages, fuzz testing technology has been widely used, and it has also improved software security to some extent \cite{DBLP:journals/tse/ManesHHCESW21}.

AFL (American Fuzzy Lop) is a leading open-source fuzz testing framework based on coverage guidance and is currently the most popular tool in this category. It offers high code coverage, strong vulnerability discovery capabilities, and operational efficiency \cite{AFL}. AFL uses instrumentation to record code coverage and employs a mutation strategy based on genetic algorithms \cite{DBLP:conf/uss/NagyNHDH21}. By evaluating the discovery of new execution paths and selecting high-quality mutation seeds, AFL reduces the generation of unnecessary test cases, increasing the fuzzing efficiency.

However, AFL has a notable limitation. Its fuzzing results are sensitive to the initial input seed \cite{DBLP:conf/issta/HerreraGMNPH21}. The quality of the initial seed determines AFL's effectiveness in covering program paths and detecting hidden vulnerabilities. If the initial seed is of poor quality or does not adhere to the program's syntax structure, AFL must conduct extensive mutation early in the fuzz testing process to pass syntax detection. This consumes significant system resources, leading to inefficient exploration of program paths and wasting fuzz testing resources and time.

To mitigate the sensitivity of AFL (American Fuzzy Lop) to initial input seeds, we propose the integration of generative adversarial networks (GANs) for seed optimization (\ourmethod{}). GANs possess the capability to discern patterns in existing datasets and synthesize new data instances from stochastic inputs. Upon completion of the training phase, the data crafted by the GAN not only mirrors the salient attributes of the source data but also captures a broad spectrum of its influential characteristics. By employing GANs, we can generate refined seed inputs that retain the diversity and complexity necessary for effective fuzz testing, thereby potentially boosting the efficacy and coverage of the AFL testing process.


Considering that GAN is prone to problems such as gradient vanishing and pattern collapse during training, which affects the quality of the output of the seed by the seed optimization model, we propose to replace GAN with WGAN, which leverages the wasserstein distance for GAN model optimization. WGAN provides stable gradient to drive the model to converge during the training process, to improve the quality of the seeds output by the seed optimization model, and to improve code coverage and the number of vulnerabilities found.

Our main contributions are summarized as follows:
\begin{itemize}
    \item We propose an AFL seed generation augmented model based on GAN. By learning from high-quality test cases, the model captures the distribution characteristics of high-quality seeds and generates higher-quality seed sets.
    \item Furthermore, We substitute GAN with WGAN, which effectively mitigates the gradient vanishing issue present in GANs. This modification enhances the training process of GAN, consequently improving the quality of generated seeds.
    \item We conduct experiments on commonly used Linux software to verify the correctness and effectiveness of the proposed method. This is achieved through a comparative analysis of code coverage and the number of vulnerabilities discovered.
\end{itemize}

The rest of this paper is organized as follows: Section II summarizes the related work of Fuzz, AFL, Fuzz optimization based on Deep Learning approaches. Section III introduces the design of our model in detail. In Section IV, we discuss the details of our experimental setup. In Section V, we give a comprehensive analysis to explain our experimental results. We summarize the advantages of our proposed work and plans for future work in Section VI.

\section{PRELIMINARY}

\subsection{Fuzzing}
Fuzz testing is a software automatic testing technology \cite{DBLP:journals/cybersec/LiZZ18}, which aims to automate the discovery of potential vulnerabilities existing in software. This method involves injecting a substantial volume of random, invalid, abnormal, or unintended data into the input data or commands of a program.

Fuzz testing fundamentally addresses a search problem within an infinite solution space \cite{liang2018fuzz}. Its objective is to identify inputs among all possible ones that can trigger program crashes, typically representing boundary scenarios or uncommon inputs. To guide fuzz testing tools in discovering such inputs, designers commonly approach the problem from two angles: generation and mutation. They integrate techniques like symbolic execution \cite{king1976symbolic}, genetic algorithm\cite{mathew2012genetic}, and other technologies to enhance the exploration capabilities of fuzz testing tools, thereby improving their code coverage and vulnerability discovery capabilities.

\begin{figure*}
    \centering
    \includegraphics[width=0.6\linewidth]{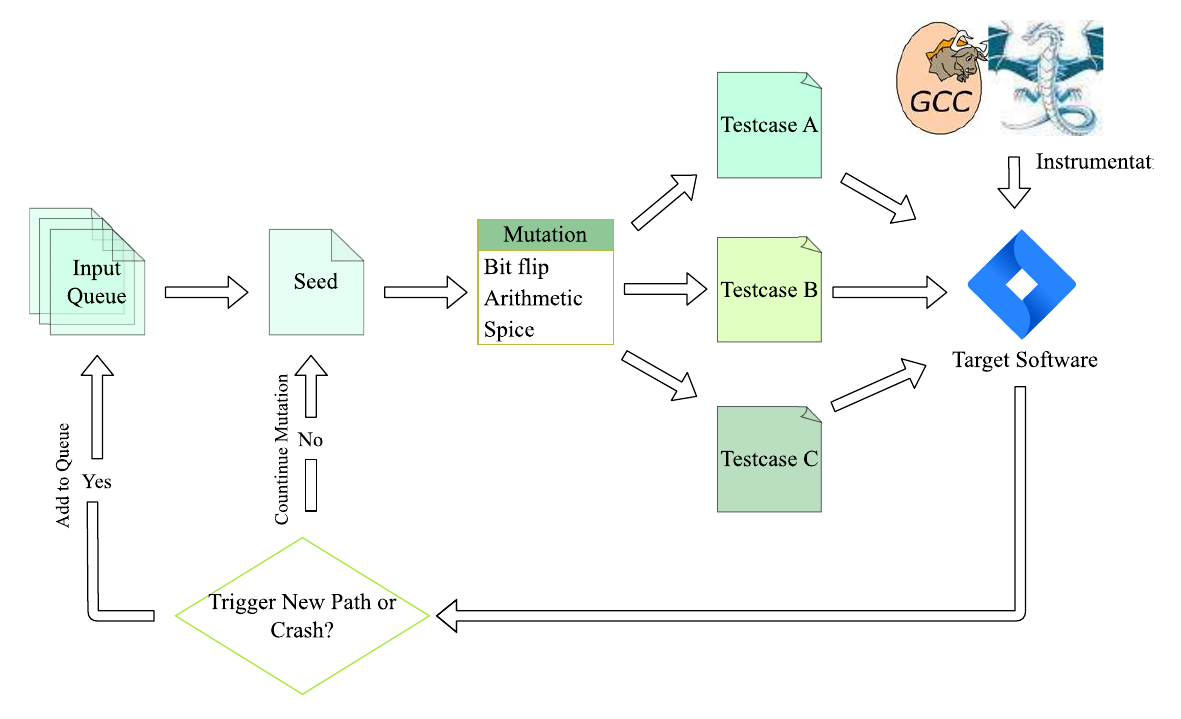}
    \caption{AFL workflow.}
    \label{AFL}
\end{figure*}

AFL is a representative evolutionary greybox fuzzer, follows the main workflow depicted in Fig.~\ref{AFL} \cite{AFL}. In each testing iteration, AFL chooses seeds from the input queue and employs various mutation strategies, including splicing and bit-flipping, to create sets of testcases. These testcases are then input into the target program, and the results are evaluated. The program's response, such as crashes or the exploration of new code paths, is detected through techniques like instrumentation and memory inspection. Based on these outcomes, the testcases are either added to or removed from the input queue.

\subsection{Generating Adversarial Network}

The generative adversarial network (GAN) is a class of machine learning frameworks introduced by Ian Goodfellow in 2014 \cite{DBLP:journals/cacm/GoodfellowPMXWO20}, consisting of two neural network models: the generator and the discriminator. The generator's objective is to produce samples that closely resemble real data, while the discriminator aims to differentiate between samples and ascertain whether they originate from genuine data distributions or are artificially generated by the generator. Throughout the training process of the GAN, the generator undergoes continuous optimization, striving to generate increasingly realistic samples to deceive the discriminator into recognizing them as originating from authentic data distributions. Simultaneously, the discriminator enhances its discriminatory capabilities to avoid being misled by the generator. This iterative optimization process continues until the generator is capable of producing samples that closely resemble real data \cite{DBLP:journals/spm/CreswellWDASB18,cdga}.

The training process of the generative adversarial network can be expressed by the loss function shown in \eqref{GAN equation}: where $x$ denotes the real data, $P_{data} (x)$ denotes the distribution that the real data obeys, $E$ denotes the mathematical expectation of the distribution, $z$ denotes the random noise generated data, and $P_z (x)$ denotes the distribution of the generated data.

\begin{equation}
\begin{split}
    \min \max V(D,G) &= E_{x \sim p_{data}(x)}[\log D(x)] \\
    &+ E_{z\sim p_z(z)}[\log (1-D(G(z))]  \label{GAN equation}
\end{split}
\end{equation}

From \eqref{GAN equation}, we can learn the training objective of a generative adversarial network. Given real data $x$ and the discriminator $D(x)$, the training goal for the discriminator is to maximize the output of $D(X)$ to approach 1. Simultaneously, for the generator function $G(z)$, where $z$ represents random noise, the training objective is to maximize the output of $D(G(Z))$ to approach $1$. The discriminator aims to correctly distinguish between real and generated data, so the training objective for generated data is to minimize the output of $D(G(z))$ to approach $0$.

\subsection{Wasserstein GAN}
GAN may encounter issues such as gradient vanishing and mode collapse during training. However, wasserstein generative adversarial network (WGAN) can mitigate the gradient vanishing problem \cite{DBLP:conf/iclr/ArjovskyB17} \cite{DBLP:conf/icml/ArjovskyCB17}, and thus we uses the WGAN as the seed optimization model instead of GAN. The wasserstein distance, as a distance metric for measuring the difference between two probability distributions, takes into account the structural and geometric features between distributions, enabling it to more accurately capture similarities and differences between distributions. Its expression is shown in \eqref{Wasserstein Distance}.

\begin{equation}
    W(P_r,P_g)=\inf_{\gamma \sim \pi (P_r,P_g)}E_{(x,y)\sim\gamma}[||x-y||] \label{Wasserstein Distance}
\end{equation}

In this expression, $\pi(p_r,p_g)$ represents the set of all joint distributions that combine $p_r$ and $p_g$, $p_r$ and $p_g$ represent the real distribution and the generated distribution, respectively. $\gamma$ represents the set of possible joint distributions, while $x$ and $y$ represent real samples and a generated sample, respectively. This paper uses the wasserstein distance as a replacement for metrics based on KL divergence or JS divergence in GAN. This approach enhances the meaningful gradients in gradient descent. By optimizing the wasserstein distance, the generated distribution $p_g$ can gradually and stably converge towards the real distribution $p_r$.

To reduce the computational difficulty of calculating the wasserstein distance, the Kantorovich-Rubinstein duality is used in GAN to approximate the calculation of the wasserstein distance \cite{ruschendorf1985wasserstein} \cite{vallender1974calculation}. The expression is shown in \eqref{Wasserstein Distance Approximate}.
\begin{equation}
    W(P_r,P_g)=\frac{1}{K}\sup_{||f||_L\leq K}E_{x\sim P_r}[f(x)]-E_{x\sim P_g}[f(x)] \label{Wasserstein Distance Approximate}
\end{equation}

In the expression, $K$ is the Lipschitz constant of the function f. After obtaining the approximate expression for the wasserstein distance, it is combined with GAN to obtain the loss functions for the generator and discriminator as shown in \eqref{LossG WGAN} and \eqref{LossD WGAN} respectively:
\begin{equation}
    Loss_G=-E_{x\sim P_g}[f_\omega(x)] \label{LossG WGAN}
\end{equation}
\begin{equation}
    Loss_D=E_{x\sim P_g}[f_\omega(x)]-E_{x\sim P_r}[f_\omega(x)] \label{LossD WGAN}
\end{equation}

\begin{figure}
    \centering
    \includegraphics[width=\linewidth]{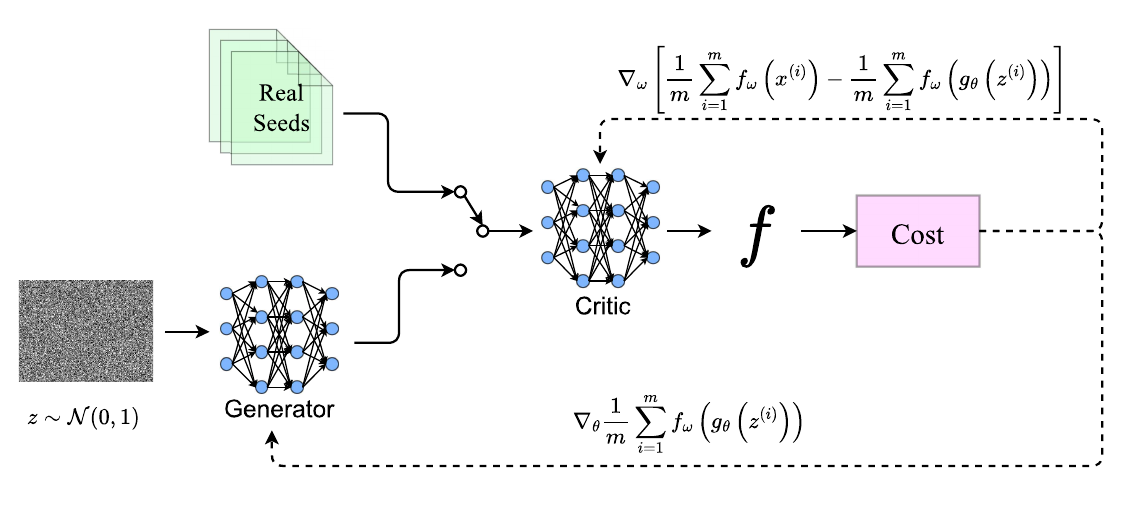}
    \caption{Working process of wgan. \cite{DBLP:conf/icml/ArjovskyCB17}}
    \label{WGAN workflow}
\end{figure}

By modifying GAN, we replace the loss function of the generator with \eqref{LossG WGAN} and the loss of the discriminator with expression \eqref{LossD WGAN}. We also replace the discriminator network with a $critic$ network. When updating the discriminator parameters, we limit their absolute values to be less than a fixed constant $c$. We also replace the original optimization algorithm Adam with the RMSProp algorithm. The algorithm flow of WGAN is as Figure. \ref{WGAN workflow}.

\section{WGAN-AFL}
\begin{figure}
    \centering
    \includegraphics[width=\linewidth]{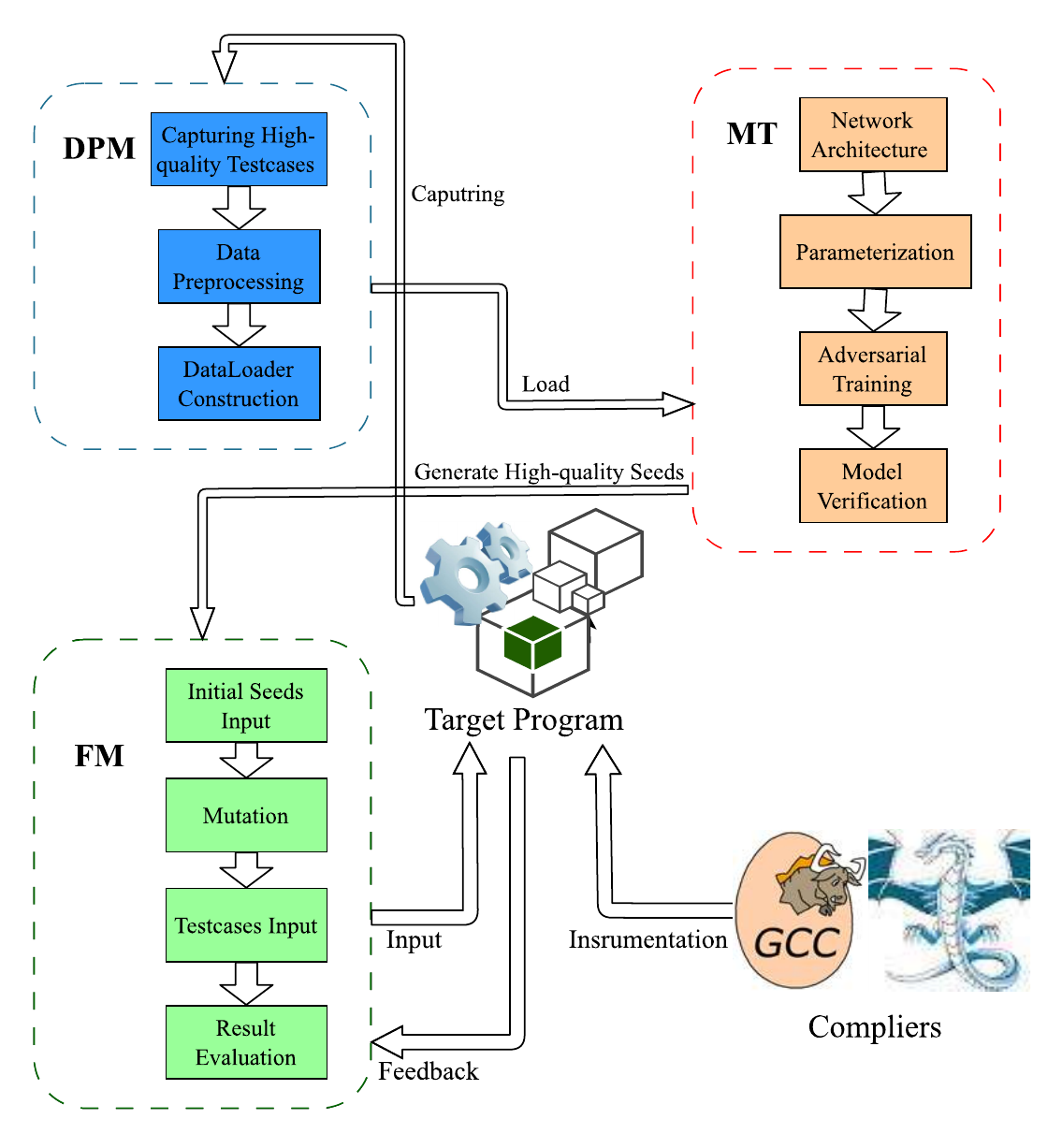}
    \caption{WGAN-AFL framework.}
    \label{WGAN-AFL framework}
\end{figure}

In this section, we describe our methodology and the main aspects of the WGAN-AFL in detail. The overall fuzzing framework is depicted in \ref{WGAN-AFL framework}, including the Data Processing Module(DPM), Model Training(MT), and Fuzzing Module(FM).

\subsection{Dataset Processing Module}
Firstly, define high-quality testcases that meets the following criteria \cite{DBLP:journals/tdsc/GanZQTLPC22}:
\begin{itemize}
    \item testcases that cause program crashes during execution;
    \item testcases with high code coverage;
    \item testcases that trigger new code paths during execution, which increase the total code coverage; 
    \item testcases with significantly different execution paths from the original input seed, which tend to trigger new execution paths;
\end{itemize}

We employ AFL to perform 72-hour fuzz testing on the software. Throughout the fuzzing process, high-quality mutation seeds are collected and filtered to serve as training data for WGAN. 

Following the capture phase, testcases must undergo preprocessing to align with the input format specifications of the neural network.  As the neural network demands tensor inputs, we initially convert each testcase in the filtered dataset into tensor form, resulting in the tensor $X$.

\begin{equation}
    X=[x_1\ x_2\ x_3 \ldots x_n] \label{tensor X}
\end{equation}

In the expression \eqref{tensor X}, $x_i$ represents the $i$th byte of the testcase $X$. As neural networks mandate uniform tensor input lengths, padding operations are necessary on each seed to ensure consistent tensor lengths. Furthermore, to improve training performance, we set the length of the padded tensor to be an integer multiple of 32. Specifically, we first determine the maximum length among the input tensors, denoted as $maxlen$, and calculate it using \eqref{maxlen} to ensure that $maxlen$ is an integer multiple of 32. Finally, we iterate through all the input tensors and pad zeros at the end of each tensor with a length less than $maxlen$ until its length is equal to $maxlen$.

\begin{equation}
    maxlen=maxlen+(32-maxlen\%32) \label{maxlen} 
\end{equation}

The above processing transforms the training samples into equally-sized tensors that can be input into the neural network for training. After completing the data initialization, the dataset can be constructed. Each element in the tensors constructed through the above operations has a size within the range of $(0,255)$. To accelerate neural network convergence, it is necessary to normalize the tensors by changing the size of each element to within the range of $(-1,1)$ using \eqref{normalize} for calculation.
\begin{equation}
    X'=(X-128)/128 \label{normalize}
\end{equation}

In this expression, $X$ represents the tensor before normalization, and $X'$ represents the tensor after normalization. After completing the normalization operation, this paper uses the dataset and dataLoader classes in the deep learning framework PyTorch to build high-quality seed sets and initialize data loaders. The data loaders are used to batch load the tensors into the neural network, improving the training efficiency of the neural network.

\subsection{Model Training}

The generator and discriminator of the WGAN implemented adopt fully connected networks. The relevant parameters are shown in Table \ref{Discriminator network in WGAN} and \ref{WGAN G}. The fully connected network structure is simple and easy to implement, requires less training time, and is able to comprehensively learn the testcases' features \cite{DBLP:journals/corr/SchwingU15}.
\begin{table}[]
    \centering
    \caption{Parameters of discriminator network in WGAN.}

    \resizebox{\linewidth}{!}{
    \begin{tabular}{cc}
    \toprule
        \textbf{Layer(type)} & \textbf{Output Shape} \\ 
    \midrule
        Fully\_connected\_1 (Linear) & (batch\_size, 256) \\
    
        LeakyReLU\_1(LeakyReLU) & (batch\_size,256)\\
    
        Fully\_connected\_2(Linear) & (batch\_size,256)\\
    
        LeakyReLU\_2(LeakyReLU) & (batch\_size,256)\\
    
        Fully\_connected\_3(Linear) & (batch\_size,1)\\
    \bottomrule
    \end{tabular}
    }
    \label{Discriminator network in WGAN}
\end{table}

\begin{table}[]
    \centering
    \caption{Parameters of generator network in WGAN.}
    \label{WGAN G}
    \resizebox{\linewidth}{!}{
    \begin{tabular}{cc}
    \toprule
        \textbf{Layer(type)} & \textbf{Output Shape} \\ 
    \midrule
        Fully\_connected\_1 (Linear) & (batch\_size, 1024) \\
    
        LeakyReLU\_1(ReLU) & (batch\_size,1024)\\
   
        Fully\_connected\_2(Linear) & (batch\_size,1024)\\
    
        LeakyReLU\_2(ReLU) & (batch\_size,1024)\\
     
        Fully\_connected\_3(Linear) & (batch\_size,1024)\\
    
        LeakyReLU\_3(ReLU) & (batch\_size,1024)\\
    
        Fully\_connected\_4(Linear) & (batch\_size,1024)\\
    
        LeakyReLU\_4(ReLU) & (batch\_size,1024)\\
    
        Fully\_connected\_5(Linear) & (batch\_size,output\_size)\\
    
        Tanh(Tanh) & (batch\_size,output\_size)\\
    \bottomrule
    \end{tabular}
    }
    \label{Generator network in WGAN}
\end{table}

The RMSProp optimization algorithm is employed for minimizing the loss function, and dynamically adjusting the learning rate to enhance the efficiency of gradient descent. Within the PyTorch framework, the optim.lr\_scheduler is utilized to systematically decrease the learning rate, facilitating a steady convergence of WGAN. Simultaneously, to prevent overfitting of the discriminator, a controlled range of noise is introduced to the real data during training, and the training labels are smoothed. This approach encourages the generator to capture the genuine features of high-quality testcases from a broader perspective.

Considering the training characteristic of WGAN, the wasserstein distance is rendered tractable by employing weight clipping in each training round, where the parameters of the discriminator are constrained to the $clip\_value$ value. Additionally, the value of $n\_critic$ is determined such that the generator updates its parameters only when the discriminator parameters are updated every $n\_critic$ iteration.

Upon convergence of the network, we preserve generator models from various rounds, and selected which are capable of producing high-quality testcases through manual validation. In the fuzzing phase, the diversity of input seeds can be enriched by sampling from these generators that output high-quality testcases.

\subsection{Fuzzing Module}
Upon convergence of the network, we preserve generator models from various rounds and select which are capable of producing high-quality testcases through manual validation. In the fuzzing phase, the diversity of input seeds can be enriched by sampling from these generators that output high-quality testcases.

With the enhanced quality of initial input seeds, the FM module's execution efficiency is further elevated, achieving higher code coverage in a shorter time. This improvement facilitates the easier triggering of software crashes, thereby enhancing vulnerability mining capabilities. Testcases that induce software crashes during testing will be preserved, enabling further analysis by developers to pinpoint and address software vulnerabilities.
\section{EVALUATION}
In this section, we evaluate WGAN-AFL's bug-finding performance and achieved code coverage with respect to the origin AFL. Specifically, we answer the following three research questions:
\begin{itemize}
    \item \textbf{RQ1.} Can WGAN-AFL find more bugs?
    \item \textbf{RQ2.} Can WGAN-AFL achieve higher code coverage and find more paths?
    \item \textbf{RQ3.} Does WGAN perform better than GAN in seed optimization?
\end{itemize}
\subsection{Experiment Setup}
In this paper, we conduct experiments on the Ubuntu 20.04-Server operating system with the specific configuration as shown in Table \ref{Experimental Environment}. The experimental subject is the commonly used toolset Binutils under the Linux system, which includes commonly used programs such as readelf, nm, and objdump, with a version of V2.25.

\begin{table}[]
    \centering
    \caption{Experimental Environment. }
    \resizebox{\linewidth }{!}{
    \begin{tabular}{cc}
    \toprule
       \textbf{Experimental Environment} &  \textbf{Configuration Information}  \\
    \midrule
        Operation System & Ubuntu 20.04 Server  \\
    
        random access memory (RAM) & 4G \\
    
        CPU & Intel(R) Gold 6133 CPU @3.0GHz \\
    
        GPU & NVIDIA Tesla T4 \\
     
        AFL Version & 2.57b \\
    
        Compilation Environment & GCC-9.3.0\\
    
        Python & 3.10\\

        Pytorch & 2.1.2\\
        
        CUDA & 11.8 \\
    \bottomrule
    \end{tabular}
    }
    \label{Experimental Environment}
\end{table}
We employing AFL-GCC for code instrumentation on the target programs in the experiment, enable the recording of code coverage throughout the testing process. This recorded information will guide the fuzz testing process based on the feedback during testing. Additionally, compared to the instrumentation mode of QEMU, source code instrumentation enhances the performance of fuzz testing and prevents unnecessary waste of system resources \cite{DBLP:conf/woot/MaierEFH20}.

To answer the research questions raised above, we design the following three sets of experiments:

\begin{itemize}
    \item \textbf{AFL group}: This is the control group that does not use any optimization method. It only uses ordinary seed inputs to conduct fuzz testing experiments with the origin AFL tool.
    \item \textbf{GAN-AFL group}: This group uses a seed optimization model based on GAN to optimize the seed inputs of AFL. GAN has the same model architecture and parameters as WGAN.
    \item \textbf{WGAN-AFL group}: This group uses a seed optimization model based on WGAN to optimize the seed inputs of AFL.
\end{itemize}

\subsection{Training Results}

\begin{figure*}
    \centering
    \includegraphics[width=0.7\textwidth]{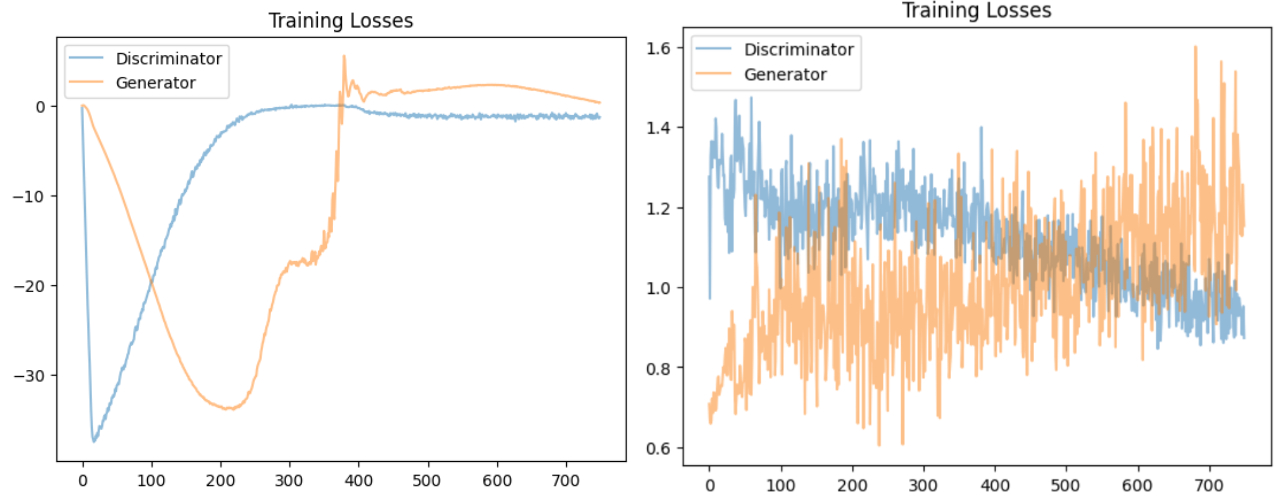}
    \caption{Variation of Loss Values during Model Training: Training Loss of GAN(Left) and Training Loss of WGAN (Right). }
    \label{Training Loss}
\end{figure*}

\begin{figure}
    \centering
    \includegraphics[width=0.8\columnwidth]{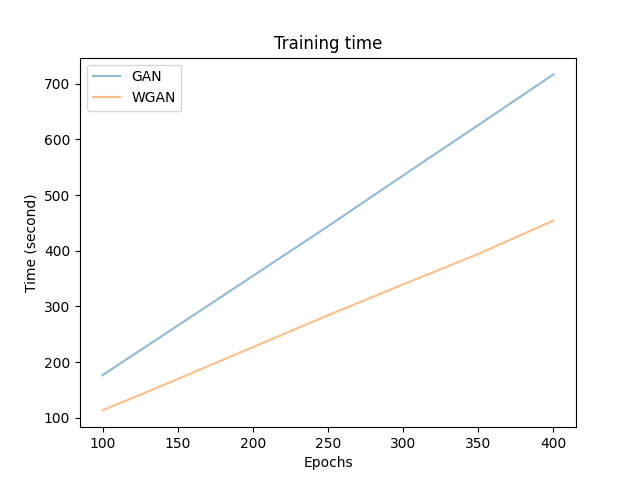}
    \caption{Training Time of GAN and WGAN. }
    \label{Training Time}
\end{figure}

Utilizing the PyTorch framework, we trained the generative adversarial network (GAN) and the Wasserstein generative adversarial network. Details about the loss variation during model training are depicted in Figure.\ref{Training Loss}, while the time consumed by model training is illustrated in Figure.\ref{Training Time}.

With epoch increases, both GAN and WGAN tend to converge, allowing the generator to grasp the features of high-quality testcases through adversarial training with the discriminator. Since fully connected networks serve as the primary model for both GAN and WGAN, the training time is relatively short, typically ranging from 5 to 8 minutes. Notably, WGAN exhibits faster training compared to GAN, attributed to a reduction in the number of gradient computations required in each round of training.

\subsection{Fuzzing Results}
An 8-hour fuzz testing experiment is conducted on four commonly used software applications (readelf, nm, objdump, and tcpdump) using AFL, WGAN-AFL, and GAN-AFL. Throughout the testing process, we recorded the code coverage and the number of discovered crashes \cite{DBLP:conf/ccs/KleesRCW018}.

\begin{table*}[]
    \centering
    \caption{Fuzzing performance of AFL, GAN-AFL, WGAN-AFL on different softwares. }
    \resizebox{0.6\textwidth}{!}{
    \begin{tabular}{cccccc}
    \toprule
    Fuzzer & Metrics & readelf & nm & objdump & tcpdump \\
    \midrule
       \multirow{3}{*}{AFL}  &  Code Coverage & 35.5\% & 25.9\% & 31.6\%& 16.4\% \\
          ~  &  New Paths & 3049 & 1203 & 868 & 1285 \\
          ~  &  Crashes & 0 & 96 & 3 & 0\\
    \midrule
       \multirow{3}{*}{GAN-AFL}  & Code Coverage & 36.4\% & 37.7\% & 33.4\% & 18.1\% \\
         ~       & New Paths & 3120 & 1926 & 917 & 1432\\
        ~        & Crashes & 0 & 208 & 19 & 0\\
    \midrule
       \multirow{3}{*}{WGAN-AFL} & Code Coverage & 41.2\% & 38.9\% & 38.3\% & 17.0\% \\
         ~       & New Paths     & 3533   & 1880   & 1052   & 1322  \\
          ~      & Crashes       &  0 & 274 & 21 & 0\\
    \bottomrule
    \end{tabular}
    }
    \label{Results}
\end{table*}

\subsubsection{Code Coverage}
In terms of code coverage, as depicted in Table \ref{Results}, WGAN-AFL exhibits the best overall performance. It achieves the highest code coverage on readelf, objdump, and nm, and the second-highest code coverage on tcpdump. The average coverage reaches 33.85\%, representing a 23.8\% average increase compared to the original AFL. This validates the superiority of the seed enhancement method based on WGAN.

GAN-AFL exhibits the second-best overall performance, but in experiments with the four sets of software, it achieves higher code coverage than the original AFL. The average code coverage reaches 31.4\%, representing a 14.8\% average increase compared to the original AFL. This indicates that GAN plays a certain role in promoting the growth of code coverage.


\subsubsection{New Paths}
In terms of new paths, WGAN-AFL also demonstrates the best performance. On readelf and nm software, WGAN-AFL exhibits a significantly higher number of program path gains compared to GAN-AFL. However, on objdump and tcpdump software, the program path gains for WGAN-AFL are slightly lower than GAN-AFL. The average number of discovered paths across the four sets of software testing reaches 1946.75, representing a 21.6\% average increase compared to AFL. This underscores the excellent expansibility of seeds generated by WGAN.

GAN-AFL also performs well, with a superior number of discovered paths compared to AFL in all four sets of software. The average number of discovered paths reaches 1848.75, representing a 15.4\% average increase compared to AFL. This validates the role of GAN in seed optimization.


\subsubsection{Crashes}
In terms of vulnerabilities detection, due to the robust security features of readelf and tcpdump, WGAN-AFL was unable to discover any existing vulnerabilities. However, in the testing of objdump and nm, WGAN-AFL demonstrated a significant advantage, uncovering the highest number of vulnerabilities among the three groups. Specifically, it discovered 274 vulnerabilities for objdump and 21 for nm. This represents a total growth of 198\% compared to AFL.

Similarly, GAN-AFL did not discover any vulnerabilities in readelf and tcpdump. However, in nm and objdump, the total number of vulnerabilities detected by GAN-AFL reached 228, representing a 130\% increase compared to AFL.


In summary, WGAN-AFL demonstrates the best overall performance, with GAN-AFL surpassing AFL. This indicates that WGAN-AFL produces seeds of the highest quality, resulting in a more diverse set of test cases during the mutation phase. These diverse test cases are more likely to explore deep program paths, thereby increasing the likelihood of triggering software crashes. While GAN-AFL generates seeds of slightly lower quality compared to WGAN-AFL, it still possesses the capability to output excellent seeds.

This indicates that WGAN exhibits the strongest learning capability, thoroughly capturing the features of high-quality test samples and minimizing the distance between the generated sample distribution and the real sample distribution. On the other hand, GAN, constrained by issues like gradient vanishing and mode collapse, can learn partial features of high-quality test samples. However, in the later stages of training, due to inappropriate distance metrics, the generator's ability to acquire effective learning gradients approaches zero or becomes overly conservative, leading to a tendency for fixed pattern outputs and a lack of sufficiently diverse seed generation.

\section{RELATED WORK}
In this section, we discuss fuzzing techniques that are based on generation and mutation. Specifically, we will explore fuzzers that leverage machine learning approaches.

\subsection{Generation-based Fuzzers}
Generation-based fuzzers focus on creating inputs systematically, often using predefined models or templates \cite{prior_template,alm,soft_template,DBLP:conf/ccs/KleesRCW018,learn_to_select,qurg,fuzzllm,white_box_llm}. They are suitable for fuzzing programs that require highly structured inputs, like interpreters, and compilers \cite{DBLP:conf/aaai/LiuLPW19}. To well utilize them, users need to provide a set of syntactic specifications and generation rules. CSmith \cite{DBLP:conf/pldi/YangCER11} hard-codes the language specification, is able to generate valid C fragment code for testing C compilers. This language specification can be defined by users, including limits on program size, the number of variables, and the depth of control flow structures. Further, Skyfire \cite{DBLP:conf/sp/WangCWL17} based on corpus and probabilistic context-sensitive grammar (PCSG), learns and deduces syntactic and semantic rules between legitimate inputs through a grammar tree, imposes appropriate heuristics and reorganization rules, and ultimately produces high-quality inputs that can pass the syntactic-semantic check. NAUTILUS \cite{aschermann2019nautilus} combines grammar rules and code coverage feedback to guide the generation of test cases. It utilizes context-free grammar to deconstruct high-quality test cases, forming a rich corpus. 


\subsection{Mutation-based Fuzzers}
Mutation-based fuzzers create a large number of testcases by applying a series of mutation rules to the seeds, and input them into the target program for fuzzing \cite{DBLP:journals/csur/ZhuWCX22,edge_case_fuzzer_llm}. AFL \cite{AFL} stands out as a representative mutation-based fuzzer. It captures the runtime code coverage of the program through code instrumentation, and subsequently employs genetic algorithms to identify valuable test samples for mutation. AFL utilizes code coverage bootstrapping to enhance both code coverage and vulnerability mining capabilities. Widely adopted by security professionals for software testing, AFL has become a prevalent tool in the field. Moreover, Some researchers \cite{DBLP:conf/acsac/ZhangHJSLLWSZL19} utilize dynamic taint analysis to trace the propagation path of data throughout program execution. This approach enables the identification of connections between potentially sensitive data and user inputs, allowing for the detection of data processing operations that may give rise to security vulnerabilities. Then, the symbolic execution method within white-box testing \cite{DBLP:conf/sp/ChaWB15} is used to identify dependencies among different input locations. 


\subsection{Machine learning for fuzzers}
The integration of machine learning has significantly enhanced the performance of fuzz testing tools. There is now a considerable number of tools leveraging machine learning methods to enhance vulnerability discovery capabilities \cite{chen2018angora}. The origin AFL is expanded \cite{DBLP:journals/corr/abs-1711-04596} by incorporating long short-term memory (LSTM) network and sequence-to-sequence (seq2seq) models \cite{fuzz_testing,xlmt,ganlm,xmt,smart_start}. These additions aim to learn the relationship between testcases and code path exploration, resulting in improved quality of variant locations within input test cases. Previous work leverages statistical machine learning \cite{DBLP:conf/kbse/GodefroidPS17} based on recurrent neural networks to automate the training of testcase generation. They selectively utilized inputs generated by the long short-term memory network model, employing a sampling strategy to achieve a balanced generation of both format-correct and format-incorrect inputs. This approach facilitates the exploration of new program states. NEUZZ \cite{DBLP:conf/sp/ShePEYRJ19} is proposed to use gradient bootstrapping techniques and the smoothing capabilities of neural networks to incrementally learn an approximation of real-world program behavior. 

\section{CONCLUSION}
We propose WGAN-AFL, an improved method to solve the problem that fuzz testing tools are sensitive to initial input seeds. Through the collection of high-quality test samples, we construct a generative adversarial network to learn their features and derive a model capable of generating high-quality initial seeds. And since generative adversarial networks suffer from training instability as well as pattern crashes, we use WGAN to alleviate this deficiency and further improve the quality of the output seeds. We use WGAN-AFL to conduct experiments on commonly used softwares under Linux, and the results show that WGAN-AFL proves to be effective due to the original AFL in terms of code coverage and the number of vulnerabilities found.
\bibliographystyle{abbrv}
\bibliography{ref.bib}

\end{document}